\documentclass[ ]{copernicus2}

\usepackage{color}
\usepackage[T1]{fontenc}

\begin{document}

\title{Auroral kilometric radiation and electron pairing
}

{\author[1,3]{R. A. Treumann}
\author[2]{W. Baumjohann$^a$}
\affil[1]{International Space Science Institute, Bern, Switzerland}
\affil[2]{Space Research Institute, Austrian Academy of Sciences, Graz, Austria}
\affil[3]{Geophysics Department, Ludwig-Maximilians-University Munich, Germany\protect\\
Correspondence to: Wolfgang.Baumjohann@oeaw.ac.at}

}

\runningtitle{AKR and pairing}

\runningauthor{R. A. Treumann \& W. Baumjohann}

\received{ }
\pubdiscuss{ } 
\revised{ }
\accepted{ }
\published{ }


\firstpage{1}

\maketitle

  

\noindent\textbf{Abstract}.-- 
{We suggest that pairing of bouncing medium-energy electrons in the auroral upward current region close to the mirror points may play a role in driving the electron cyclotron maser instability to generate an escaping narrow band fine structure in the auroral kilometric radiation. We treat this mechanism in the gyrotron approximation, for simplicity using the extreme case of a weakly relativistic Dirac distribution instead the more realistic anisotropic J\"uttner distribution. Promising estimates of bandwidth, frequency drift and spatial location are given.   } 

\introduction

The plasma dynamics in the near-Earth high-altitude auroral magnetosphere is comparably easy to monitor, either from ground or space \citep[e.g.,][]{paschmann2003}. The peculiarity of this region lies in the overlap of the dense, cold, electrically neutral atmosphere and the dilute, fully ionized, collisionless, hot magnetospheric plasma which occurs in a narrow layer of roughly 100-300 km ($\sim0.03$ R$_E$) vertical extent. It forms the topside sufficiently electrically resistive ionosphere to, at its bottom, allow cross-magnetic field currents to close the magnetic field-aligned currents that connect the distant magnetosphere and Earth. Continuously growing evidence suggests their origin in reconnection in the central-magnetotail cross-field current-sheet. From a terrestrial viewpoint the currents flow either upward or downward. Depending on the wave or particle picture, they are carried by (kinetic) Alfv\'en waves or electrons. 

In the particle picture electrons of energies $E_e$ in the range of several keV emerge from the reconnection sites to flow downward along the magnetic field into the ionosphere, carrying upward currents. Correspondingly, downward currents are carried by  low-energy, $E_e\lesssim$ few 0.1 keV, upward accelerated ionospheric electrons. This picture is well established and was strongly supported by observations from the Viking \citep{lundin1987,deferaudy1987},
Freja \citep[cf.,][]{lundin1994}, and FAST  \citep[see the special issue on FAST, introduced by][]{carlson1998} spacecraft. 

Among the processes related to field-aligned currents, generation of  Auroral Kilometric Radiation (AKR) \citep{gurnett1974} and its fine-structure still remains insufficiently understood. AKR is broadband and intense, propagating in the free space magneto-ionic X mode.  It releases several per cent of the total auroral energy available during magnetospheric disturbances into free space, an amount substantially exceeding any reasonable gyro-synchrotron radiation above the X or O mode cut-offs. 
AKR is non-thermal and highly variable. Its mechanism has been convincingly traced back to direct amplification of free space radio waves, mainly in the X mode, by the Electron Cyclotron Maser instability (ECMI)\citep{sprangle1977a,sprangle1977b,wu1979,melrose1985,melrose2008,cairns2005,bingham2013,speirs2005,speirs2008,speirs2014,treumann2006,mcconville2008}, a possibility based on negative absorption (adopted from maser theory). This was first identified by \citet{twiss1958} and \citet{schneider1959} as a possibility to convert a plasma from an absorber into a radiator, while not yet referring to the electron cyclotron instability. It requires an underdense background plasma immersed into the strong auroral magnetic field $\mathbf{B}$. The ratio of cyclotron to plasma frequency $\omega_c/\omega_e>1$ must be large for its generation and escape. Physically this means that there are not enough electrons available in the volume to digest the radiation. Thus the non-absorbed part is radiated away into space. 

The responsible field-aligned electron component responsible for instability and radiation must be uplifted into a metastable elevated energy state for the collective absorption coefficient to change sign, allowing the plasma volume as a whole to radiate quasi-coherently like masers or lasers in the infrared respectively optical bands. The currently favoured condition resembling this kind of excitation is the electron loss-cone distribution, lacking electrons in a narrow field-aligned angular domain. They mimic the necessary excited state. The free energy stored in the loss cone is directly transferred to the free space modes by the ECMI.  

Attempts of the plasma to refill the loss-cone by low frequency resonant wave-particle interaction with VLF \citep{labelle2002} are  slow quasilinear processes under the dilute topside conditions.  They effectively limit the high-energy radiation belt fluxes  \citep{kennel1966} but cannot come up for either depleting the lower energy auroral field-aligned electron loss cone or explaining the variability and fine structure of AKR, though models have been put forward \citep{louarn1996} to overcome this deficiency. 

Other mechanisms have also been proposed, based on electron-hole dynamics \citep{treumann2011}; for their nonlinearity they did not find general acceptance. Though the existence of  holes has been observationally confirmed \citep{ergun1998a,ergun1998b,pottelette1999,pottelette2002}, and the electron hole mechanism nicely reproduces several properties of the radiation, it lacks confirmation. Holes are mesoscopic features few Debye lengths long only.  It is difficult to see how they could effectively amplify km-lengths waves. This might still be possible statistically as holes exist in very large numbers organized in chains along the magnetic field. They might act collectively over the AKR kilometer wavelength to amplify radiation, as has been suggested \citep{treumann2011}. Such a detailed stochastic calculation is still missing.

In the following we propose, at least qualitatively, another promising mechanism possibly capable of causing the spatial and temporal fine structure observed in AKR. This mechanism can be based on the resonant interaction of quasi-trapped electrons with propagating plasma waves generating an attraction between electrons spaced by a Debye length along the field. 

\begin{figure*}[t!]
\centerline{\includegraphics[width=0.65\textwidth,clip=]{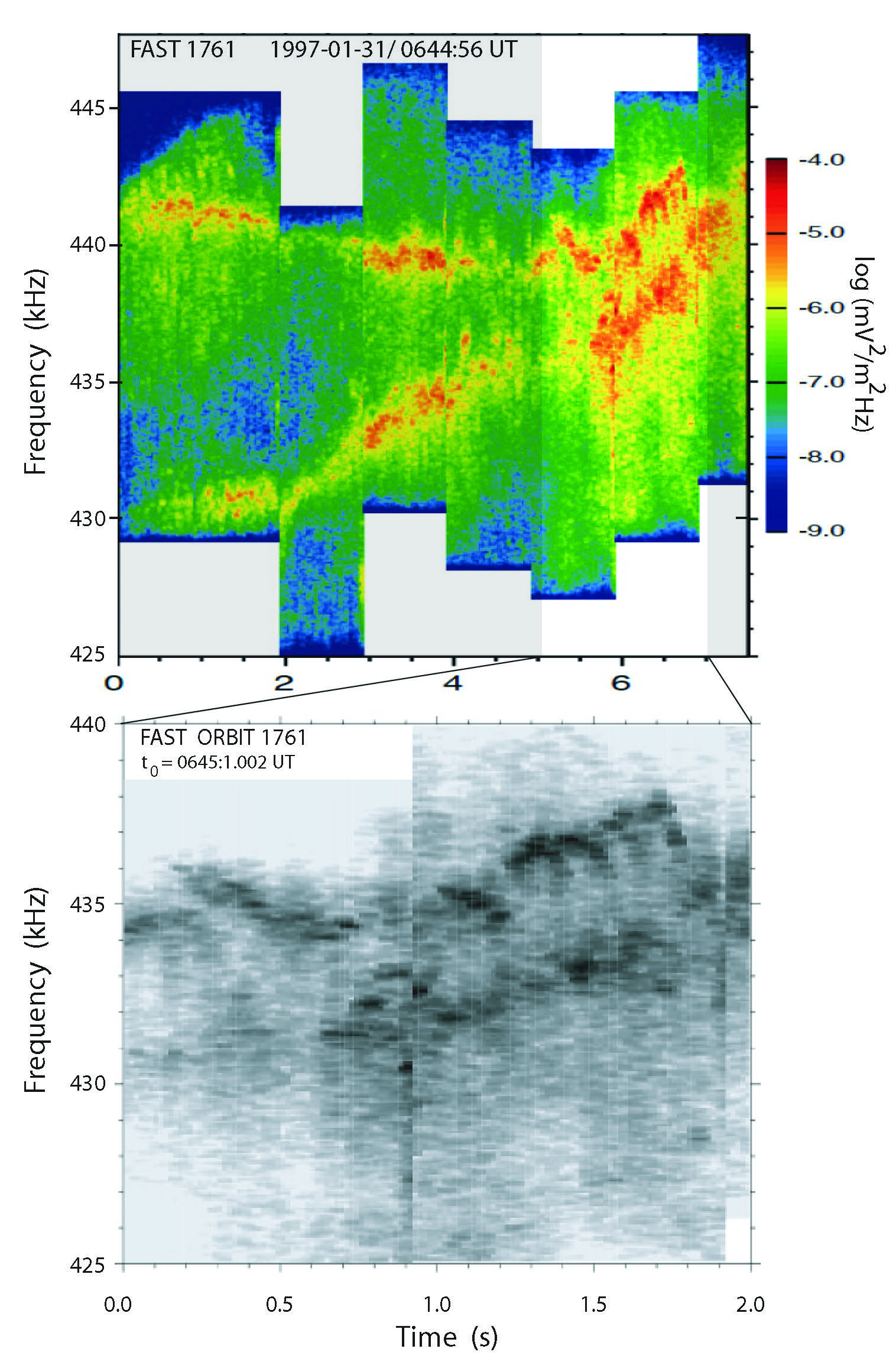}}
\caption{{Fine structure in auroral kilometric radiation FAST observations \citep[Figure adopted from][]{treumann2011}. Top: two narrow bands of intense and highly temporally structured drifting auroral kilometric emissions. The bandwidth is of the order of a few kHz only. Upward drift of the bands implies downward motion of the source into the stronger ambient magnetic field region. Obviously each band consists of many separate short emission events. Bottom: High temporal resolution of the indicated available time interval clearly showing the superposition of the many microscopic radiation events which make up the two radiation bands. Each event moves first upward in frequency (sources moves downward in space), then turns around and moves back upward. Emission is strongest in the turn-around (reflection).} } \label{fig1}
\end{figure*}

\section{Topside electron pairing}
 
Field aligned electrons move in the geomagnetic mirror geometry. Downward current-upward electrons of  low ionospheric energy, starting at large pitch-angles in the upper ionosphere, when propagating outward into the magnetosphere, conserve their magnetic moments and become quickly completely field aligned. Upward current-downward electrons on the other hand, starting at small pitch-angles in their low-field reconnection site, while moving along their separatrices, increase their pitch-angle. Some of them may ultimately become trapped in the magnetospheric field if only their mirror points lie above the ionosphere. 

Reconnection in the tail current sheet provides two electron populations, an almost strictly field-aligned population escaping along the separatrix, and an exhaust component that impacts on the reconnection-caused plasmoids and is scattered into larger pitch-angles. These latter electrons remain magnetospherically trapped along the auroral magnetic field with mirror points remaining well above the ionospheric current closure.   

Recently we have shown that, in a magnetic mirror geometry, conditions can evolve when classical electron pair-formation occurs. This process generates a very high thermal (pitch-angle) anisotropy of the paired component. Conditions of this kind evolve in mirror modes but are also expected in the upward current region. Below we qualitatively develop this scenario to some detail as a preparation to its quantitative investigation via a more elaborate analytical theory of its properties, or via numerical simulations.    

\subsection{Attractive electron potential}
Classical electron pairing is based on collective over-screening of the bare electron charge outside the Debye sphere  of moving electrons in resonance with a plasma wave of frequency $\omega_k$. The parallel electric potential $\Phi$ caused by an electron of velocity $\mathbf{v}$ is obtained from
\begin{equation}
\Phi(\mathbf{x},t)=-\frac{e}{(2\pi)^3\epsilon_0}\int d\omega d\mathbf{k}\frac{\delta(\omega-\mathbf{k\cdot v})}{k^2\epsilon(\mathbf{k},\omega)}e^{i\mathbf{k}\cdot(\mathbf{x}-\mathbf{v}t)}
\end{equation}
where the dielectric function contains the susceptibilities $\xi_{i,e}$ contributed by the waves and the Debye screening term $(k\lambda_D)^{-2}$. Note that the time variation of any electromagnetic potential $\mathbf{A}(\mathbf{x},t)$ does not contribute. Since the dielectric appears in the denominator its inverse comes into play. Quite generally it can be written
\begin{equation}\label{eq-inv-resp}
\frac{1}{\epsilon(\omega,\mathbf{k})}=\frac{k^2\lambda_D^2}{1+k^2\lambda_D^2}\Big(1+\frac{\omega_{\mathbf{k}}^2}{\omega^2-\omega_{\mathbf{k}}^2}\Big)
\end{equation}
where $\omega_k$ is the solution of the dielectric $\epsilon(k,\omega)=k^2c^2/\omega^2$, considering only the plasma eigenmodes. The potential depends on the resonant electron contribution with the eigenmodes. 

 {The inverse dielectric suggests that, in particular cases, the potential can become negative, if only theplasma wave frequency $\omega_k^2\gtrsim\omega^2$, in which case the second term in the brackets becomes negative and larger than one. Once this happens, the Debye term is superseded by the resonance. The contribution to the above electric potential $\Phi$ then is negative, such that $\Phi>$-integral changes sign. Screening then results in attraction outside the Debye radius $x>\lambda_D$. This corresponds to an overscreening of the electron charge in resonance $\omega/k\approx v$ with the particle speed which happens in the region behind the particle. What is thus needed for this to happen is that the electron speed is comparable to the phase velocity $v\sim\omega/k$, i.e. resonance.\footnote{{It is interesting to note that a similar effect would also result in purely growing/damped waves where $\omega_k=\pm i\gamma_k$, an effect not yet explored anywhere.}}.} 

For ion-sound waves we derived the precise conditions \citep{treumann2019}. The physics of attractive potential generation is the same for kinetic Alfv\'en waves (kAW) but the conditions are slightly different \citep{narita2020b}. We refer to the latter paper {for details even though they are not as important for the purposes of the present communication}. For the kinetic Alfv\'en wave dielectric one has 
\begin{equation}
\epsilon_{kAW}(\omega,\mathbf{k})=1+\frac{1}{k^2\lambda_D^2}+\frac{c^2}{V_A^2}{\bigg(1+k_\perp^2\lambda_e^2\bigg)}\bigg[1+\big(\mathbf{k}\cdot\mathbf{r}_{ci}\big)^2\bigg(\frac{3}{4}+\frac{T_e}{T_i}\bigg)\bigg]^{-1}
\end{equation}
with $\mathbf{r}_{ci}=\mathbf{v}_{i\perp}/\omega_{ci}$ the vectorial ion gyro-radius{, and $\lambda_e=c/\omega_e$ the electron inertial length. The relevant wave frequency applicable to the strong field auroral region for $\beta=2\mu_0NT_e/B^2\lesssim \sqrt{m_e/m_i}$ is 
\begin{equation}
\omega_{\mathbf{k}A}^2\approx k^2V_A^2/\sqrt{1+k_\perp^2\lambda_e^2}
\end{equation}
with $V_A\ll c$ the Alfv\'en speed of the ordinary Alfv\'en wave. The correction factor in brackets, though it can easily be included, plays no essential role for our purposes.} There is a vast literature on the kAW \citep[cf. ][for the most recent collection of their properties in the solar wind where the plasma $\beta>1$, the other limiting case]{narita2020a}. It is long-wavelength and, in a hot plasma, propagates somewhat faster along the magnetic field than the ordinary Alfv\'en wave, and perpendicular to it roughly several ten times slower. In cold plasma where electron inertia cannot be neglected it is instead slower than the Alfv\'en wave. For our purposes here it suffices to know that the kAW  propagates approximately at Alfv\'en speed and, in the topside AKR source slows somewhat down. Here the ion contribution plays no role, and one has approximately $V_{A}\to V_A/\sqrt{1+k^2\lambda_e^2}$. The weak wave electric potential \citep{lysak1996} resulting from its kinetic nature is along the magnetic field and is believed to be responsible for some electron acceleration, sometimes (in the older literature) claimed to come up for the whole auroral electron energy. In view of the tail-reconnection set-up this seems improbable; the electrons start from there with  energy in the right range.  Any attractive pairing effect in resonance with those fast parallel electrons will be in this direction as well but is independent on the wave field as pairing is a single particle effect which causes a local change in the dielectric constant of the plasma but otherwise has no effect on the wave Alfv\'en dynamics. Here we explore its importance for electron dynamics and generation of AKR. Naturally, like any Alfv\'en wave, the kinetic Alfv\'en wave possesses a perpendicular electric field which under certain conditions  causes the electrons to perform a  perpendicular drift motion which displaces the electrons very slowly from their original to a neighbouring flux tube, an effect which we safely ignore below.
\begin{figure*}[t!]
\centerline{\includegraphics[width=0.75\textwidth,height=0.4\textheight,clip=]{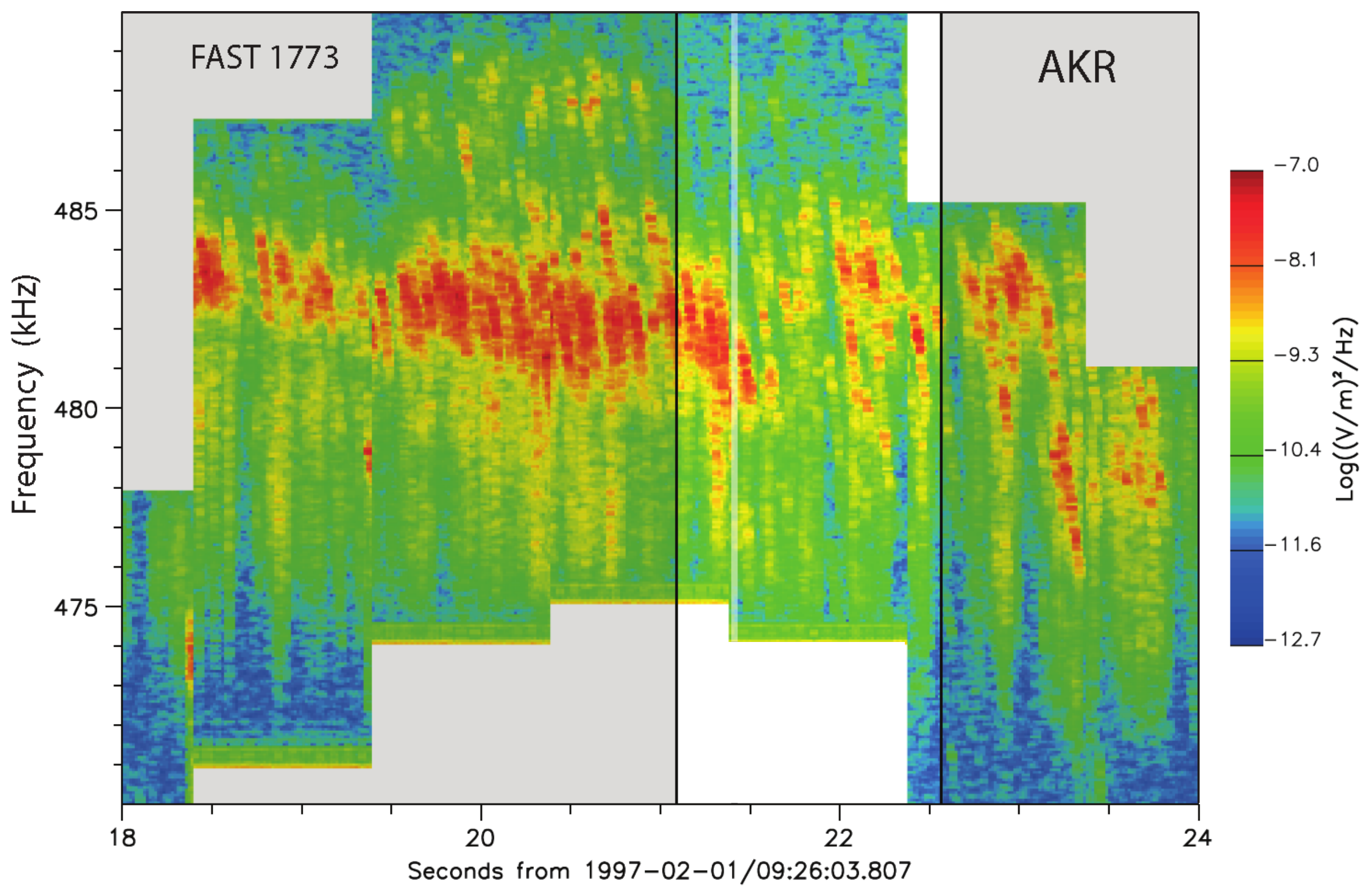}}
\caption{{Another example of narrow band emission in auroral kilometric radiation \citep[after][]{treumann2011} observed by FAST with no high-resolution observations available. In this event time resolution does not permit resolving all microscopic emission sources. Only few spatially downward (upward in frequency) moving sources can be identified. The emission band is about stably in spatial (and frequency) location. Apparently most moving microscopic radiation sources move upward in space (downward in frequency) from their location of strongest emission.}} \label{fig2}
\end{figure*}

\subsection{Pairing scenario}
Wave-electron resonance between the electron and kAW produces a negative potential in the electron wake which  traps another electron, a process that modifies the plasma dielectric. Attractive potentials were first suggested half a century ago by \citet{neufeld1955}\footnote{\citet{neufeld1955} introduced it  even before Cooper's invention of electron pairs \citep{cooper1956,bardeen1957} of opposite spins, the idea of which Cooper might have gotten from there \citep[and another paper by][who did not succeed]{frohlich1952} and applied successfully to cold fermions.} who, however, did not suggest any pairing mechanism. They considered Langmuir waves which, however, turned out to have no importance anywhere. The idea was later picked up in application to ion-sound waves  \citep{nambu1985}. In fact, in an un-prepared plasma, attractive potentials and pairing are indeed unimportant, except of possibly justifying the assumption of compound formation, as is assumed in pic simulations in plasma, \citet[see][]{treumann2014}. 

In particular cases attractive potentials may play a role in dilute magnetized plasmas when electrons in magnetic mirror geometries undergo bounce motions, conditions occurring for instance in mirror instabilities,\citep{treumann2019}. For the description and precise calculation of the mechanism causing classical electron pairing, one may consult that paper. Pairing in plasma is based on the evolution of the attractive potential in the wake of a moving electron and can, in principle be extended to multiple electrons. Classically,  at the applied high temperatures the paired electrons have of course for long lost their fermionic property. 

Since the wake is  of much longer scale than the Debye length $\lambda_D=v_e/\omega_e$, with $mv_e^2=2T_e$ the classical thermal speed of the electron at temperature $T_e$, and $ \omega_e$ the electron plasma frequency, the attractive potential evolves \emph{outside} the electron Debye sphere. It unavoidably affects many electrons which are separated at distance $N^{-1/3}\ll\lambda_D$, with $N$ background plasma density. Thus there will always be at least another electron which moves at approximately same speed and thus feels the attractive potential of the first which, however becomes modified by its presence, with modification rapidly decreasing, the more electrons participate. 

The condition that the trapping is about stable is that the parallel energy difference between the two electrons is $u_\|^2<2|e\Phi|/m$, where $u_\|$ is defined below. We have shown that the basic effect is that two electrons bind together in the attractive potentials which overcompensates the remaining repulsive potential outside the Debye sphere. It requires the presence of a low frequency plasma wave propagating parallel to the two interacting electrons and is in resonance. The resonance is a single particle effect and no plasma resonance. The main condition is that the parallel phase velocity $\omega_k/k_\|$ of the wave and the parallel centre-of-mass velocity 
\begin{equation}
U_\|= \frac{1}{2}(v_{1\|}+v_{2\|})
\end{equation}
of the two electrons ${1,2}$ which fall into the narrow interval $\lambda_D<\Delta s\lesssim1.5\lambda_D$ outside the Debye sphere are about the same. This implies that for the two coupled electrons holds
\begin{equation}
u_\|=\frac{1}{2}|v_{1\|}-v_{2\|}|\ll |U_\|| 
\end{equation}
The centre of mass velocity of the electrons parallel to the field in the bounce motion would decrease to zero when, after having started from some location $s(0)$ along the field, they approaching the common mirror point $s_m$. 

In paired resonance with the kAW the resonance stops the bounce motion as long as the particles remain to be paired. The pair of electrons then continues moving as a triplet together with the wave at the slow wave speed $\omega/k_\|\approx V_A$ along the field. Thus, when the attraction comes into play, the parallel speed of the pair drops to that of the wave.

Assume that the electrons initially each had high energy $E_e$ and thus high parallel speed $U_\|$. Close to their mirror point just before pairing their energy is about completely transferred to their perpendicular speeds $v_\perp$, i.e. the gyration. At short distance from $s_m$ the parallel speed $U_\|$ has decrease so much that it equals the phase velocity $\omega_k/k_\|$ of the kAW, and the attractive potential evolves. The newly formed pair becomes locked to the plasma wave by the induced local change in the dielectric function which is responsible for the attractive potential. The triplet of the two electrons and the plasma wave then continues moving together at the slow phase speed, and the paired electrons completely drop out from  bounce, both maintaining {their} perpendicular speeds. Conservation of the magnetic moments, which affects only the perpendicular energy implies that the latter increases. Thus pairing causes adiabatic perpendicular heating of the pairs. At this point their parallel speed has become
\begin{equation}
U_\|(s\lesssim s_m)\sim\omega/k_\|\ll U_\|(0)
\end{equation}
much less than their initial parallel velocity $U_\|(0)$.

When this happens, the perpendicular individual electron speeds equal their initial velocities $v_{\perp 1,2}(s_m)\approx v_{1,2}(s=0)$. Here $s$ is the distance along the magnetic field from the reconnection to the pairing site, which is close to the initial mirror point $s_m$. Note that the trapping condition is that electrons 1 and 2 have about same parallel speed, which is given by the above condition on $u_\|$.  It fixes the parallel speed, locks the electrons to the wave, and thus, by conservation of the electron magnetic moments $\mu=\mathcal{E}_\perp(s)/B(s)$, transfers all their remaining individual energies into their perpendicular speeds. Since the cyclotron frequency depends only on the magnetic field $B(s)$ at that spatial location, this process  modifies the gyroradii of the two electrons. 

Assume that the perpendicular speed is roughly about their average thermal speed $v_e$, then their perpendicular energy in pairing becomes comparable to their initial energy (temperature $T_b$) which is the mean energy of the bouncing plasma component. It is high above that of the local ambient plasma if any. The error made in this assumption is small, because the energy spread of the fast electrons generated in tail-reconnection must necessarily be small. The pitch-angle spread of the field-aligned electrons in the topside auroral region is reduced to a few degrees only, filtering out a narrow range of energy spread available for pairing. Those electrons are close to mono-energetic and possess a large energy (thermal) anisotropy which can roughly be approximated as 
\begin{equation}
A_p=\frac{\mathcal{E}_\perp(s_m)}{\mathcal{E}_\|(s_m)} -1\equiv \tan^2\theta(s_m)-1\approx \frac{2\mathcal{E}_e(0)}{m u_\|^2} \gg 1
\end{equation}
where the parallel energy is $mu_\|^2/2$, and in bounce motion 
\begin{equation}
\tan^2\theta(s_m)=\lim_{s\to s_m} \Big[\frac{B(s_m)}{B(s)}-1\Big]^{-1} \gg 1
\end{equation}
 is very large. A more precise calculation is to be based on a combination of the bounce motion and the pairing condition. 
 
Considering the smallness of $u_\|$, this anisotropy is huge. As written here, it is an energy ratio. However, in a volume of many Debye lengths along the magnetic field in which a large fraction of electrons, i.e. a large fraction of electrons in the Debye sphere contribute to trapping, each carrying along its paired partner electron at a distance of $\lambda_D+\Delta s$ while propagating down the field along with the kAW. There will be a substantial fraction of such pairs in the flux tube over one or several wavelengths of the kAW. These may not be stable for very long time due to fluctuations but new pairs will continuously reform within the unceasing stream of electrons supplied by tail reconnection. Such a volume provides a fairly large energetic anisotropy and, under not too restrictive conditions, may drive the ECMI unstable.   

Electrons, after acceleration and ejection from the reconnection site (the reconnection exhaust, as it is sometimes called) have nominal velocity $v_0\gtrsim 10^4$ km/s. Their parallel speed is $v_\|=v_0\cos\theta(s)$. The phase velocity of kAW is of the order of $\omega/k\sim 10^3$ km/s. When the electron approaches resonance, the pitch-angle has changed to satisfy the condition
\begin{equation}
\cos\theta(s)\sim \frac{\omega}{kv_0}\sim \frac{V_A}{v_0}\lesssim 10^{-1}
\end{equation}
which close to its mirror point $s_m$ corresponds to a pitch-angle $\theta(s)\gtrsim 85^\circ$. Its perpendicular speed is thus practically $v_0$, and the anisotropy in energy $A\gtrsim 10^2$, as argued above. If contributed by a susceptible number of particles an anisotropy that high must have a profound effect on the generation of radiation. 

Total plasma densities in the broad upward current AKR source region amount to at most  $N\sim$ a few times $10^6$ m$^{-3}$, corresponding to a dilute and even underdense strongly magnetized plasma whose cyclotron frequency is around $\omega_c/2\pi \approx 300$ kHz corresponding to a magnetic field of  $B\lesssim 10^4$ nT ($\lesssim 0.1$ Gauss). This density is mostly due to the presence of  downflow electrons. The Debye length is of the order $\lambda_D\sim 1$ m.  The mean inter-particle distance is $d\sim 10^{-2}$ m. This gives roughly 150 particles within a length of $1.5\, \lambda_D$. Let the number density of pairs be $N_p$, then a probably  realistic pair-to-plasma density ratio would be $N_p/N\sim 10^{-3}$ or so, which means that just every thousandth electron would capture another one to form a pair. The remaining electrons form the plasma background and do  not contribute to any anisotropy. They might, however, be subject to a weak loss-cone distribution of the kind of a Dory-Guest-Harris (DGH) or Ashour-Abdalla-Thorne (AAT) distribution. Background and paired electron populations are completely independent. They couple only by quasi-neutrality $N=N_0+N_p$.

Below we examine, for simplicity and to demonstrate the possible excitation of radiation, just one particular well-known emission model which may apply to the distribution of  paired electrons. This is the gyroresonant (gyrotron) scenario.

\subsection{Gyroresonant Emission} 

The simplest imaginable radiator of electromagnetic waves in a highly anisotropic plasma is the gyrotron. Radiation is due to electron bunching in the unstable free-space radiation wave field of frequency $\omega$ and wave number $\mathbf{k}$, not the kAW field! This wave field obeys the electromagnetic dispersion relation $\mathcal{N}\equiv k^2c^2/\omega^2\approx 1$ sufficiently far above X-mode cut-off. It implies the presence of a highly anisotropic particle distribution capable of directly amplifying one of the free-space magneto-ionic radiation modes. Its effect is a maser emission driven by the anisotropy.  It has been suggested originally for plasma devices, gyrotrons \citep{gaponov1959,gaponov1967}. The highly anisotropic electron (pair) distribution in the parallel moving frame of the kAW is then modelled as a gyrotropic Dirac distribution with all the (relativistic) momentum $\mathbf{p}_\perp$  in the perpendicular direction
 \begin{equation}
f_p(p_\|,p_\perp)= \frac{N_p}{2\pi p_{0\perp}}\delta(p_\perp-p_{0\perp})\delta(p_\|-m{\gamma}\omega_k/k_\|)
\end{equation}
neglecting here the small velocity spread $u_\|$ in the parallel direction and any possible spread in the distribution of perpendicular momentum $p_{0\perp}$ due to any given initial electron distribution, in order to make it analytically accessible. Including such a spread would imply the use of an anisotropic J\"uttner distribution \citep{treumann2016} which complicates the problem substantially. For our purposes it suffices to stick to the simplest model first. In the paired state the electrons move with wave phase velocity $V_A\ll c$ along the magnetic field. We therefore transform to the wave frame $p_{\|A}$ setting $\delta(p_\|-mV_A)\to\delta(p_{\|A})$. {[Actually, this transformation is not as simple because the relativistic factor $\gamma$ also depends on perpendicular momentum, a complication which we neglect here as the small Alfv\'en velocities .] } Then, in the laboratory frame, any displacement of the pair along the spatially changing (increasing or decreasing) magnetic field $B(s)$ appears, in the observed pair-caused emission spectrum, simply as a frequency drift.  The perpendicular particle moment must however be treated relativistically with initial relativistic factor $\gamma_0=\sqrt{1+ p_{0\perp}^2/m^2c^2}$. Moreover one assumes that $k_\perp\rho\approx 0$ the perpendicular wavelength of the emitted radiation is large with respect to the relativistic electron gyro-radius $\rho$. In this case, in the Bessel expansion of the plasma dielectric \citep[cf., e.g.,][]{baumjohann1996} only the harmonics $n=0,\pm1$ survive, and the dispersion relation of the free space modes reads \citep{melrose1985}
\begin{equation}
\mathcal{N}^2\equiv\frac{k^2c^2}{\omega^2}= 1-\frac{\omega_p^2}{\omega^2}\Big[\frac{\omega}{\omega-\omega_c}+\frac{p^2_{0\perp}}{2m^2c^2}\frac{k^2c^2-\omega^2}{(\omega-\omega_c)^2}\Big]
\end{equation}
The last term in the parentheses is the relativistic correction which turns out to be crucial. The growing solutions of this relation with positive imaginary part $\omega_i>0$ of the frequency corresponds to oblique propagation. The maximum growth rate becomes
\begin{equation}
\frac{\omega_i}{\omega_c}\approx \sqrt{3}\Big[\frac{p_{0\perp}^2}{4m^2c^2}\frac{\omega_p^2}{\omega_c^2}\Big(1-\mathcal{N}^2\cos^2\theta\Big)\Big]^\frac{1}{3}
\end{equation}
with $\mathcal{N}\approx 1$ and $1-\mathcal{N}^2\cos^2\theta\approx\sin^2\theta$, under the condition $\omega^2_p/\omega^2_c\ll \sin^4\theta\,(p_{0\perp}/mc)^4$ \citep{melrose1985}. One hence requires that the emission is very close to perpendicular such that $\sin^2\theta\approx 1$. One also needs a strong magnetic field $B$ and low density $N_p$ of the pairs to have $\omega_p\ll\omega_c$, and the initial momentum, respectively the relativistic $\beta= v_{0\perp}/c$ of the electrons, should not be too small. For $\sim 10$ keV {electrons one} has roughly $\beta\approx 0.1$. Emission at the fundamental in its turn {then} requires  $\omega_p/\omega_c< 10^{-3} -10^{-2}$. This is not unreasonable in view of the rough order-of-magnitude discussion given above. Estimated densities would approximately correspond to this condition. One, however, expects that even under these conditions the presence of the non-paired plasma background would absorb radiation at the fundamental, such that the intensity of the radiation in the fundamental should become low.

The present calculation has been done just for the fundamental harmonic $|n|=1$. Radiation at higher harmonics is much less vulnerable to absorption in the diluted plasma of the AKR source region.  There can be no doubt that higher harmonics $|n|>1$ will be excited as well, though possibly at weaker than nominal fundamental growth rate at $|n|=1$,  while escaping re-absorption. Unfortunately inclusion of higher harmonics becomes substantially more complicated \citep{melrose1985} analytically because all the different Bessel functions of higher order $|n|>1$ appear in the dispersion relation of the radiation modes, and one must consider the full dielectric tensor. Even with the simple Dirac distribution the problem then becomes almost intractable. We may, however, for the purpose of this perspective letter, put forward an approximate argument, just to infer what the dominant higher harmonic effect would become. With inclusion of low higher harmonics, the relativistic resonance condition (frequency mismatch) reads 
\begin{equation}
\Delta_n = \omega -n\omega_c/\gamma -k_\| v_\|
\end{equation}
The harmonic number just factorizes the relativistic cyclotron frequency. In order to infer the effect of $|n|>1$ to lowest approximation, one may replace the cyclotron frequency in the expression for the growth rate with $n\omega_c$ to obtain that 
\begin{equation}
\frac{\omega_{i,n}}{n\omega_c}\sim \Big[\frac{p_{0\perp}^2}{4m^2c^2}\frac{\omega_p^2}{n^2\omega_c^2}\Big(1-\mathcal{N}^2\cos^2\theta\Big)\Big]^\frac{1}{3}
\end{equation}
One therefore expects the growth rate of the low higher harmonics weakly decreases as $\sim n^{-2/3}$ when normalised to its $n$th cyclotron harmonic, which implies that at the lowest harmonics the growth increases with respect to the fundamental increases as $\sim n^{1/3}$. In addition to the fundamental with its problem of escaping, reabsorption, and quasilinear quenching when remaining trapped, those low harmonics may indeed dominate the observation. Realizing this fact is important as it affects the interpretation concerning the involved magnetic fields. 

The above result, though still quite imprecise, can also be interpreted as $n^{-2}\sin^2\theta$ which means that, holding up the conditions on the density, \emph{the emission at higher harmonics is  more oblique}. In addition \citep[see][]{melrose1985} the frequency mismatch $\Delta_n$ must be positive in order to escape from the plasma. This is anyway necessary but easier to satisfy for higher harmonics than the fundamental. $\Delta_n>0$ puts the emission band above the $n$th harmonic, and the condition on the parameters becomes $\omega_p/n\omega_c\ll\beta^2\sin^2\theta$ which is slightly less restrictive on the density, while relaxing the escape condition for all harmonics under consideration which are above the X-mode cut-off. Thus the expectation is that one observes pair-excited higher harmonics rather than fundamental radiation as the latter will be suppressed.

\section{Discussion}
Above we tentatively applied the idea of electron pair formation due to the generation of attractive potentials between electrons in bounce resonance with kinetic Alfv\'en waves in the auroral region. In the simplest gyrotron model the resulting weakly relativistic dilute pair population transforms the upward current region into a gyrotron which works due to the comparably high energy (temperature) anisotropy of the pair population. The expectation in this case is that part of the auroral kilometric radiation is emitted in gyroradiation at some harmonics. In this respect one may note that harmonic radiation has indeed been observed though not analyzed in detail. More important is, however, that temporarily highly variable structures have regularly been  detected in the high resolution observations of FAST \citep{pottelette1999,pottelette2002,treumann2011} both in the upward and downward current regions. 

The conditions for sole gyrotron emission still seem to be severe. They become relaxed for moderately high harmonics. On the other hand, the gyrotron Dirac distribution may not be ideally chosen to describe the real situation. It should thus be taken just as a first idealized step to a physical interpretation of the fine structures which adds to some former models \citep{louarn1996,treumann2011} from which they substantially differ  while not replacing them completely as each of them has its advantages. 

The gyrotron is an extreme model. It however shows that a paired population may indeed contribute to radiation, and one might consider its consequences when applying the ECMI theory to a combination of a loss cone and a pair population. So far only loss cones \citep[][and others]{dory1965,ashour1977,wu1979,louarn1996,melrose2008} have been taken into account for the reason that they are natural distributions in mirror geometries. They become efficient when accounting for the relativistic effect. The same applies to the gyrotron emission proposed here. So a combination of both is quite promising. This will be deferred to a separate investigation. In the next subsection we describe some ideas concerning the fine structure of AKR in view of gyrotron radiation.  

\subsection{AKR fine structure} 
For data sets on AKR fine structure the reader may consult our above cited previous publications on this matter. We have included some of them here in Figures 1 and 2.  AKR is by no means a structureless banded emission close to the local gyrofrequency as suggested by any of the emission models. High temporal and frequency resolution of its upward current region \citep[{for the full upward current data set see}][{their Fig. 3, not included here}]{treumann2011}  shows that the radiation  consists of at least two components: a relatively broadband rather weak and quasi-stationary emission spectrum on which a large number of intense drifting narrow-band emissions is superimposed {as shown in the top of Fig. \ref{fig1}}. These may drift up and down in frequency at various frequency drifts. In many cases high temporal resolution {(Fig. \ref{fig1}, bottom)} shows these little structures to move up and down in frequency and even to turn around or vanish at a certain place in the spectrum, often with the most intense radiation emitted just in the turn-around. We have tentatively in previous work attributed this kind of motion to the presence of electron holes which are known to drift up and down along the magnetic field. However, in view of the above problems with the hole emission model we attempt to address the gyrotron model to these structures. 

Though it is by no means clear that the background component is indeed homogeneous -- the highest resolution case available to us seems to indicate that it is simply the unresolved overlap of many drifting fine structures emitting radiation at larger distance from the spacecraft (FAST). But consider just the most intense banded radiation {(Figs. \ref{fig1} and Fig. \ref{fig2}  in particular the $t\sim 2$ s long high-temporal resolution part in Fig. \ref{fig1} )}. 

The upward (positive) drift on the upward leg cannot be well resolved even at the given high time resolution. It is about vertical amounting roughly to $\Delta \omega/2\pi\Delta t >10/ 0.1= 100$ kHz/s. The total emission band is restricted in this case to the interval $425<\omega/2\pi<440$ kHz. In the topside auroral region the curvature of the magnetic field is weak over a change in cyclotron frequency of this magnitude. The emission band consists of $\Delta t\lesssim 0.3$ s long steeply drifting emissions which start from lower frequency, move up in frequency until reaching its climax where the radiation intensifies, and after passage turn around to low frequencies again under substantial radiative softening. Such emissions are typically V-shaped. 

The 0.1 s time the radiation is above detection threshold implies that, at nominal Alfv\'en speed of $V_A\sim 1000$ km/s, the paired electrons have moved not much more than a distance of $\Delta s\approx 100$ km along the magnetic field together with the Alfv\'en wave. This means that they have been very close to their mirror point. 

One may  speculate that the turn-around in the emission where the intensity of radiation maximizes in a narrow emission band of not more than $\Delta \omega/2\pi\sim2-3$ kHz bandwidth indicates that the group of radiating paired electrons has passed their mirror point soon leaving the carrier wave, returning to bounce motion and dissolving while injecting their excess energy into radiation. The return is indicated by the turn-around during which the drift of the emission briefly turns negative from high to low frequency, i.e. from higher to lower magnetic fields. The high radiation intensity at turn-around frequency is probably due to slowing down of the large perpendicular momentum electrons in this volume to roughly zero parallel velocity, which happens when leaving the kAW. This necessarily leads to a high volume emissivity.

The fact that there are several such intense narrow emission bands which by themselves drift across the spectrum indicates that in this model several groups of paired electrons should exist with independent dynamics and possibly occupying different magnetic flux tubes. 

{Figure \ref{fig2} is a particularly interesting case.} No high resolution data were available during this period. Here, the steep nearly unresolved positive drifts are accompanied by weak emissions while the turn-arounds and substantially flatter negative drifts show intense emission when the electrons move at much slower speed when slowly picking up the bounce speed. Of course, in this interpretation no directivity of the emission is included as this has not been measured. 

In any case, the most intense emission is in the turn-around, a narrow band of bandwidth typically less than $1$ kHz. Let us assume that the emission is at the fundamental $n=1$. 

The different groups of electrons, spaced in frequency at their turn-arounds have sightly different mirror points and thus probably have slightly different initial pitch-angles $\alpha_0$ or are in resonance with a  different kAW. The latter is probable if the slopes of the emission bands at turn-around are different and they occupy the same flux tube. In the second half of Fig. 1  the turn-arounds are parallel but at slightly different frequencies separated by $\sim2$ kHz. Assuming a dipole field, the center frequency of 430 kHz {corresponds to} a magnetic field of $B(R)\approx 15360$ nT or $B/B_0\approx 0.44$. This in the dipole field implies an altitude of roughly $h_{n=1}\approx 2013$ km above Earth's surface. And the $\sim 1$ kHz bandwidth implies that the turn-around AKR source of maximum intensity in fine structure has a vertical spatial extension of $\Delta h\lesssim 15$ km, where in all these estimates we neglected the latitude dependence of the magnetic field. 

Each of the V-shaped emissions in such a model then corresponds to a separate group of paired electrons moving along in one flux tube with their resonant kAW. These groups of pairs follow about regularly for a while every $\delta t\gtrsim 0.4$s. If this is the period of the kAW, its wavelength is of the order of $\lambda\gtrsim 2500$ km, not an unreasonable value. It is intriguing that such a periodicity exists which otherwise is not easy to understand, unless it is attributed to the tail reconnection injection mechanism.

These estimates apply to emission at the fundamental $|n|=1$. When this is suppressed say either by the escape condition with frequency mismatch $\Delta<0$ or by absorption in the background electron population, then emission of AKR will occur at the second harmonic $|n|=2$ or higher. In this case the location of the AKR source has to be replaced to a magnetic field of the order of $B\approx 8000$ nT which is at an altitude $h_{n=2}\approx 2540$ km above Earth, slightly larger than for emission at the fundamental.

\subsection{Summary}
Sporadic very intense emission of electromagnetic waves, in particular in the radio band, as is frequently observed in solar astrophysics and sometimes also from remote astronomical objects, gives a clue to the understanding of the internal physics of the emitting regions. It thus is useful as a convenient remote probe. In basic electrodynamics such radiation relies on the simple gyro-synchrotron mechanism which principally, because it is of higher order, is weak and, in order to become intense, requires very large systems to increase the emission measure. Such systems evolve usually very slowly such that sporadic emissions can hardly come up for any short term intense variations like, for instance, the recently discovered and now quite frequently observed extremely short broadband radio bursts. 

Moreover, at low energies particle scattering in the sources do not contribute; they are spared for the much higher energy range of X rays. Therefore emissions are sought for which are capable of causing intense fast sporadic radiation in the radio regime different from synchrotron emission, the favoured mechanism in astrophysics since it is so simple. The electron cyclotron maser instability is probably the most promising in magnetized media. However, it requires particular conditions set in the source which much be satisfied. The most important is that the radiating plasma does not reabsorb the emission, i.e. its absorption coefficient must become either very small or negative. The latter case is realized in the ECMI when the emitting electron population is lifted into an excited state and at the same time there not much background plasma is available to reabsorb the radiation. The latter is most easily realized at higher harmonics since radiation at the fundamental is mostly damped due to trapping, quasilinear quenching, and reabsorption. 

In the present letter we have attacked the problem of generation of the pronounced fine structure superimposed on auroral kilometric radiation (AKR) from the terrestrial topside ionosphere (or near-Earth magnetosphere) under disturbed auroral conditions. Generation of such fine structures in the emission is a non-trivial problem considering that the emission band is just a few kHz wide, which is at most 1\% of the bandwidth of AKR. It is believed that the latter is caused by the general weakly relativistic pitch-angle distribution of auroral electrons, a theory first proposed by \citet{wu1979} which accounts for the relativistic deformation of the resonance between electrons and free space modes through the inclusion of the dependence of the resonance on the transverse electron velocity $\beta_\perp$ appearing in the relativistic $\gamma$ factor. Inversion of the absorption property of the rather dilute plasma in this case is caused {at the expense} of the perpendicular (gyrational) energy of the plasma while still requiring the loss-cone as the agent of providing the demand of free energy stored in the inverted occupation of higher energy levels. Since it is quite difficult to believe that the most intense and extremely narrow-band AKR emission is excited by the global loss cone distribution, we have attempted another mechanism. Such a mechanism can possibly be found in the generation of an attracting (negative) electrostatic potential $\Phi$ in the wake of moving electrons as proposed long ago \citep{neufeld1955,nambu1985} and recently found \citep{treumann2019} to be applicable to magnetic mirror geometries in space plasma. The attractive potentiantial is a single particle effect which locally modifies the plasma dielectric  just outside the Debye sphere of the particle in cooperation (resonance) with a plasma wave moving along the magnetic field and parallel to the electron. It requires that the phase velocity and the parallel speed of the electron match, $\omega_k/k\approx v_\|$. Since the phase velocity is low, for bouncing electrons this condition implies that the effect occurs just close to the mirror point of the electron where the electron slows down to approach the phase speed of the wave. This happens naturally for the auroral high-energy electron component in resonance with kAWs. The attractive potential than traps another electron and locks both electrons to the phase speed of the wave, a process in which many electrons should be involved over the wavelength of the wave along the magnetic field. It naturally produces a highly-anisotropic electron distribution which we modelled as a displaced Dirac distribution, typical for gyrotrons \citep{gaponov1959}, which is a simplified weakly relativistic version of the ECMI.  We applied the gyrotron theory to the auroral conditions, proposing that it may explain the observation of narrow-banded intense AKR emissions probably rather at the harmonics than the fundamental. A more elaborate theory requires inclusion of the background plasma and relaxing the Dirac distribution to become a relativistic particularly suited Maxwellian \citep[anisotropic J\"uttner, see][]{treumann2016} distribution. This, however, lies outside the purpose of the present perspective note.

The proposed mechanism looks promising in application to the auroral zone. It might also have other astrophysical relevance, for instance in view of narrow band non drifting solar radio bursts like Type I emissions. Another possible application would be to the mysterious extremely short broadband cosmic radio bursts. Since the gyrotron mechanism, if sufficiently intense, is capable of simultaneously generating a large number of harmonics it could possibly be responsible for their emission in not too strong magnetic field configurations such that the harmonic emission bands are not separated too far in frequency.

\section*{Acknowledgments}
We acknowledge valuable discussions with the late Johannes Geiss,  R. Nakamura, Y. Narita, and P. Zarka. RT acknowledges the contributions of R. Pottelette to all earlier work on the subject of auroral kilometric radiation.

\end{document}